\documentclass[fleqn,usenatbib]{mnras}

\usepackage[T1]{fontenc}
\usepackage{ae,aecompl}
\usepackage{courier}

\usepackage{graphicx}	
\usepackage{amsmath}	
\usepackage{amssymb}	
\usepackage{booktabs,array,tabularx}
\usepackage{epstopdf}

\makeatletter 
\def\eqalign#1{\null\,\vcenter{\openup\jot\m@th  \ialign{\strut\hfil$\displaystyle{##}$&$\displaystyle{{}##}$\hfil      \crcr#1\crcr}}\,} 
\makeatother

\DeclareMathAlphabet{\mathsc}{OT1}{cmr}{m}{sc}
\def\testbx{bx}%
\DeclareRobustCommand{\ion}[2]{%
\relax\ifmmode
\ifx\testbx\f@series
{\mathbf{#1\,\mathsc{#2}}}\else
{\mathrm{#1\,\mathsc{#2}}}\fi
\else\textup{#1\,{\mdseries\textsc{#2}}}%
\fi}

\title[Modelling carbon and oxygen in the quiet Sun]{The influence of photo-induced processes and charge transfer on carbon and oxygen in the lower solar atmosphere}

\author[R.P. Dufresne et al.]{
R.P. Dufresne,$^{1}$\thanks{E-mail: rpd21@cam.ac.uk}
G. Del Zanna,$^{1}$
and N.R. Badnell$^{2}$
\\
$^{1}$DAMTP, University of Cambridge, Wilberforce Road, Cambridge CB3 0WA, UK\\
$^{2}$Department of Physics, University of Strathclyde, Glasgow G4 0NG, UK
}

\date{Accepted XXX. Received YYY; in original form ZZZ}

\pubyear{2020}

\begin{document}
\label{firstpage}
\pagerange{\pageref{firstpage}--\pageref{lastpage}}
\maketitle

\begin{abstract}
To predict line emission in the solar atmosphere requires models which are fundamentally different depending on whether the emission is from the chromosphere or the corona. At some point between the two regions, there must be a change between the two modelling regimes. Recent extensions to the coronal modelling for carbon and oxygen lines in the solar transition region have shown improvements in the emission of singly- and doubly-charged ions, along with Li-like ions. However, discrepancies still remain, particularly for singly-charged ions and intercombination lines. The aim of this work is to explore additional atomic processes that could further alter the charge state distribution and the level populations within ions, in order to resolve some of the discrepancies. To this end, excitation and ionisation caused by both the radiation field and by atom-ion collisions have been included, along with recombination through charge transfer. The modelling is carried out using conditions which would be present in the quiet Sun, which allows an assessment of the part atomic processes play in changing coronal modelling, separately from dynamic and transient events taking place in the plasma. The effect the processes have on the fractional ion populations are presented, as well as the change in level populations brought about by the new excitation mechanisms. Contribution functions of selected lines from low charge states are also shown, to demonstrate the extent to which line emission in the lower atmosphere could be affected by the new modelling.
\end{abstract}




\section{Introduction}
\label{sec:intro}

The transition region in the solar atmosphere bridges the lower temperature, higher density chromosphere at the base with the tenuous, high temperature plasma in the corona. The transition region (TR) undergoes a change in temperature from around 25000\,K at the bottom to about 600000\,K at the interface with the corona. Furthermore, these changes take place in a region only a few thousand kilometres thick. For predicting chromospheric line emission nothing short of time-dependent modelling with radiative transfer and (magneto-)hydrodynamics, along with atomic models featuring multiple levels, electron collisional, atomic collisional and photo-induced processes, will suffice. This is demonstrated by such works as \cite{kerr2019}, \cite{lin2017} and \cite{judge2003}. At the other end of the scale, much coronal diagnostic work may be carried out using the independent atom model, which separates the calculation of internal level populations for each ion from determining the fractional populations of the ions. Furthermore, it primarily considers electron collisional and radiative decay processes alone. Thus, in the coronal approximation, ion populations may be determined by transitions between the ground levels of each ion with no regard for the internal structure of the atom. The method is employed, for example, in the \textsc{Chianti} atomic database \citep{dere1997}. The procedure works well because the presence of more highly ionised charge states and low densities in the corona means many ions are in the ground level. Consequently, numerous ion balances have been produced over the years, such as those by \cite{jordan1969}, \cite{arnaud1985} and \cite{mazzotta1998}. 

For those who wish to use spectroscopic techniques to diagnose the properties of the TR, this produces a dichotomy over which method to use. At some point in the solar atmosphere, the simple coronal modelling will break down and more sophisticated modelling should be employed. Factors which may alter atomic processes, or new processes not previously considered, have to be incorporated, depending on the conditions in the region of the solar atmosphere being modelled. \cite{burgess1969} show that as densities increase dielectronic recombination (DR) is reduced and the charge state distribution is correspondingly altered, even at coronal densities. \cite{nussbaumer1975} subsequently confirmed this for carbon in the TR, but also demonstrate how diagnostics for \ion{C}{ii} would be altered by photo-ionisation. In coronal modelling this process is typically ignored, and the much-cited reference text, \textit{The Solar Transition Region}, by \cite{mariska1992} states that photo-ionisation has no influence on the diagnostics of the quiet solar atmosphere. The claim of \cite{nussbaumer1975} regarding photo-ionisation, though, has been confirmed by \cite{rathore2015}. Furthermore, \cite{rathore2015} show that, when chromospheric modelling is applied to carbon, \ion{C}{ii} formation takes place in the chromosphere; the coronal modelling of \textsc{Chianti} \citep{dere1997} predicts that its formation would be in the TR. Just as importantly, \cite{rathore2015} show how line ratios from this ion are significantly altered by optical depth effects. 

Another process known to alter the temperatures and densities at which ion stages form is charge transfer, which involves the exchange of electrons during collisions between atoms and ions. The ionisation potential is substantially lower for this process compared to that for collisions with free electrons, and it is more frequently adopted in other areas of astrophysics where lower temperatures prevail than in solar physics. Because it is a resonant process between neutral oxygen and hydrogen, it is, however, routinely employed for modelling emission from \ion{O}{i} in the chromosphere. \cite{baliunas1980} also show that the process substantially alters the silicon ionisation equilibrium at TR temperatures, affecting all charge states up to \ion{Si}{iv}. Despite this, the effect has been less explored for other ions and elements in the solar atmosphere.

Attention has been placed in more recent years towards improving the coronal approximation for modelling higher density regions. \cite{nikolic2013, nikolic2018} produced analytical fits to estimate the reduction in DR rates which \cite{burgess1969} and \cite{summers1974} demonstrate. A further alteration in the charge state distribution as density increases is shown by \cite{nussbaumer1975} and \cite{summers1983} to arise from ionisation and recombination from long-lived, metastable levels. To facilitate modelling of the effect of metastable levels, data has been produced by, for example, \cite{ludlow2008} and \cite{ballance2009} for electron impact ionisation (EII) from excited levels, and \cite{badnell2003} and \cite{badnell2006} for recombination. \cite{dufresne2019} and \cite{dufresne2020} recently applied these techniques to collisions involving free electrons to investigate how the diagnostics for carbon and oxygen would change in the TR, and found that the emission from carbon and oxygen ions which form lower down in the TR were improved by 20-40 per cent, whereas the majority of oxygen lines which form higher up in the TR were little changed in comparison to the coronal approximation. It was noted, however, that there are still considerable discrepancies in the predicted intensities of lines from lower temperature ions when compared to observations.

The aim of this work, then, is to explore further atomic processes that may affect the charge state distribution in the solar transition region. The focus is photo-induced processes and charge transfer, which have already been shown to alter the charge states of some elements lower down in the TR, as discussed above. These effects have been incorporated into the models for carbon and oxygen built by \cite{dufresne2019} and \cite{dufresne2020}, which already include the other effects described above. The next section details the reasons for including these atomic processes and the methods and data sources used to include them. Section\;\ref{sec:results} shows how the new modelling affects predictions for ion formation, level populations and line contribution functions compared to the coronal approximation. A short conclusion will be given in Sect.\;\ref{sec:concl}.

\section{Methods}
\label{sec:methods}

Photo-induced processes are independent of electron density, and are commonly included in more tenuous plasma, such as the solar corona and nebul\ae. Photo-ionisation (PI) processes may be important in the high density regions of the chromosphere and transition region because they are close to the photosphere, and so its strong emission will be less diffuse. The low charge states of the main elements present in the Sun are multi-electron systems, meaning their cross sections will be larger than the more highly charged coronal ions. The ionisation potentials are also smaller for these ions, and the radiation is orders of magnitude stronger at those wavelengths than at the ionisation thresholds of coronal ions. Photo-excitation (PE) could also have an effect in the lower atmosphere because of the cooler temperatures. Where excited levels in the low charge states are more than a few electron-volts (eV, 1\,eV\,=\,11604\,K) higher than the ground level, electron impact excitation (EIE) rates will be relatively small. By contrast, the solar flux at these energy separations is at its strongest, making PE a potentially important excitation mechanism, which could enhance the populations of upper levels and alter line emission. Stimulated emission, on the other hand, is less likely to have the same impact on total de-excitation rates because it is added to the spontaneous decay rates, which are also independent of density and temperature.

Charge transfer (CT) reactions occur in collisions between atoms and ions when one or more electrons are exchanged between the reactants. The abundances of hydrogen and helium in most astrophysical settings mean that only they are considered as perturbers. In the case of a charge transfer reaction between an element A with charge $+z$ and hydrogen, for example, it occurs in the following way,

\begin{equation}
 A^{+z}~+~H~\rightarrow~A^{+(z-1)}~+~H^+~.
 \label{eqn:ct}
\end{equation}

\noindent The reverse reaction is also possible, depending on the nature of the target and the collision dynamics, and is known as charge transfer ionisation. The threshold for CT will be the difference in the individual ionisation potentials of $A^{+(z-1)}$ and H. In the case of CT between neutral oxygen and hydrogen, then, the threshold is 0.02\,eV, which is much smaller than the ionisation potential of neutral oxygen when it is involved in collisions with free electrons (13.62\,eV). CT ionisation for atomic oxygen could, therefore, take place at much lower thermal energies than ionisation by electron impact.

In this work, the modelling will be treated in ionisation equilibrium, which helps assess the importance of the atomic processes in their own right, separate from dynamical effects taking place in the atmosphere. Because ionisation equilibrium is assumed here, the models will be run with parameters that best reflect conditions in the quiet Sun.

\subsection{Atomic processes}
\label{sec:atommethods}

\subsubsection{Electron collisional processes}
\label{sec:elecmethods}

The same collisional-radiative (CR) models for carbon and oxygen used in \cite{dufresne2019} and \cite{dufresne2020} were used in this work. They include ionisation from metastable levels and suppression of DR. A few changes were made to the data incorporated into the models. For \ion{O}{ii} the electron impact excitation data at low temperatures between the ground and metastable levels from \cite{tayal2007}, already in \textsc{Chianti}, were retained, but the new data from \cite{mao2020n} were incorporated for all other transitions. For \ion{O}{iii} the EIE rate coefficients were taken from \cite{mao2020c}. Recently, \cite{morisset2020} note that discrepancies are found when using the \ion{O}{iii} data from \cite{mao2020c}. This, however, is for optical lines formed below 25000\,K in nebul\ae. To circumvent this problem, the data from \cite{storey2014} for the ground configuration at low temperatures were retained, and Mao et al. used for all other transitions. More details on this may be found in the new release of \textsc{Chianti} v.10 \citep{delzanna2021}. To check the new EIE rate coefficients, predicted intensities were calculated using the same models and observations in \cite{dufresne2020}. Compared to the results from that work, the predicted intensities using the new data showed changes of no more than 20 per cent. This highlights that different EIE rate coefficients cannot account for the discrepancies found in Dufresne et al. for the \ion{O}{ii} lines and the \ion{O}{iii} intercombination line.

To accommodate the new excitation process with neutral hydrogen, the \ion{O}{i} model in $LS$-coupling from \cite{dufresne2020}, using EIE rates from \cite{tayal2016b} and radiative decay rates from \cite{tachiev2002} and \textsc{Autostructure} \citep{badnell2011}, was supplemented with four more terms. These are the triplet and quintet terms arising from the $2p^3\;4d,\,4f$ configurations, so that the bound terms match those given by \cite{barklem2007}. EIE data from \cite{barklem2007} was included for all transitions involving the new terms; radiative decay rates involving the terms were incorporated from \textsc{Autostructure}. Electron impact ionisation and photo-ionisation rates from approximations were added for the excited levels above metastable in ions which were seen to be affected by the new excitation processes. The approximation of \cite{vriens1980} was used for EII and the hydrogenic approximation of \cite{karzas1961} for PI. The only exception was in \ion{O}{i}, where PI cross sections above metastable were obtained from \textsc{Autostructure}. Three-body recombination rates into the included levels were checked and found to be far below other process at these densities. The only effect not included which may alter the final results for the level populations is level-resolved recombination into excited states.

\subsubsection{Photo-induced processes}
\label{sec:pimethods}

The same methods used by \cite{dufresne2019} to calculate photo-ionisation rates for carbon were carried out here for oxygen. The PI rate for an incident photon of frequency $\nu$ absorbed by an ion in a bound level $i$ and ionising to a final level $j$ in the next higher charge state is given by

\begin{equation}
\alpha^{PI}_{ij} = 4 \pi \; \int\limits_{\nu_0}^\infty \frac{{\sigma_{ij}(\nu)}} {h\nu} \; J_\nu \; {\rm d}\nu \, ,
\label{eqn:pirate}
\end{equation}

\noindent where $\nu_0$ is the threshold frequency below which the bound-free cross section $\sigma_{ij}(\nu)$ for the process is zero. Here,

\begin{equation}
J_{\nu} = \frac{\Delta \Omega} {4 \pi} \;  \overline{I}_\nu = 
W(r) \; \overline{I}_\nu \, ,
\label{eqn:dilution}
\end{equation}

\noindent where $\Omega$ is solid angle, $W(r)$ is the dilution factor of the radiation, that is, the geometrical factor which accounts for the weakening of the radiation field at a distance $r$ from the Sun, and $\overline{I}_\nu$ is the average disc radiance at frequency $\nu$.

\begin{figure*}
 \centering
 \includegraphics{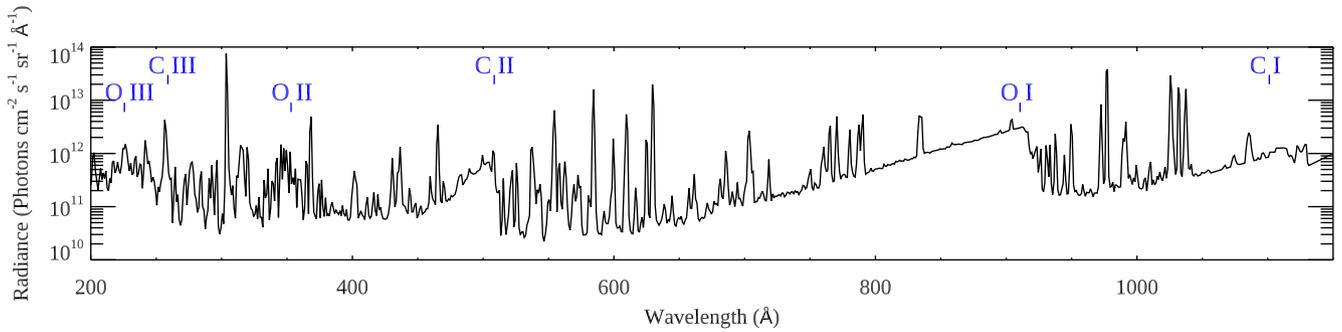}
 \caption{Radiances derived from the Whole Heliospheric Interval reference spectrum of Woods et al., which was used for the photo-induced processes. The thresholds for photo-ionisation of the low charge ions are also indicated.}
 \label{fig:solarflux}
\end{figure*}

For the spectral radiance of a field expressed in terms of wavelength, the PE rate from a lower level $i$ to upper level $j$ is given by

\begin{equation}
 \centering
 P_{ij}~=~\frac{g_j}{g_i}\frac{\lambda^5_{ji}}{8\pi hc}W(r)U_\lambda A_{ji},
 \label{eqn:perate}
\end{equation}

\noindent where $A_{ji}$ is the spontaneous radiative decay rate for the corresponding downward transition between the two levels, $g_i$ is the statistical weight of level $i$, and the energy density is given by $U_\lambda = \frac{4\pi \overline{I}_\lambda}{c}$. The rate for stimulated emission is the same expression, except it is independent of the statistical weights. The rates were added to the models for every transition for which there was a radiative decay rate.

For consistency with regards to detailed balance with radiative recombination, the PI cross sections from the radiative recombination (RR) work of \cite{badnell2006} were used for oxygen, as made available on the Atomic Processes for Astrophysical Plasmas (APAP) website\footnote{www.apap-network.org}. The average disc radiance is obtained from the Whole Heliospheric Interval (WHI) reference spectrum of \cite{woods2009}. While this is an irradiance spectrum, and converting it to radiances is not always straightforward, it has the advantage of covering a wide spectral range. The results were compared to the more highly resolved radiances of the Solar Ultraviolet Measurements of Emitted Radiation (SUMER) instrument on board the Solar and Heliospheric Observatory (SOHO), and were found to be in good agreement. For all transitions at wavelengths longer than the WHI spectrum (24000\,\AA) a blackbody spectrum was used with a thermal temperature of 6100\,K. A portion of the spectrum is shown in Fig.\;\ref{fig:solarflux}, which also shows the threshold wavelengths for photo-ionisation of the low-charge ions. An uncertainty in the method for calculating rates for photo-induced processes is how much lines emitted in one region of the atmosphere will affect ions in another part. The main emission mostly comes from the photosphere and chromosphere below, which the ions considered here should experience. The majority of the remaining emission affecting these ions will be from ultra-violet (UV) lines, which is largely emitted in the same region as that in which the ions form. A proper treatment using radiative transfer would provide greater accuracy, but is beyond the scope of this paper. A constant dilution factor of 0.45 was used for the TR, which corresponds to halfway up the solar TR, at approximately 3000\,km from the photosphere. (The dilution factor changes little from the bottom to the top of the TR.)

\subsubsection{Charge transfer}
\label{sec:ctmethods}

Early estimates of CT cross sections and rate coefficients relied on the Landau-Zener approximation, which is more suitable for higher collision energies because of the assumptions made. Considering the low collision energies in many of the astrophysical situations in which charge transfer will be relevant, the rate coefficients derived using this approximation may have considerable inaccuracy. Consequently, newer results derived using quantum mechanical methods have been sought when including the process in the models. In the cases where rate coefficients have not been provided or where a wide enough temperature range is not given, rate coefficients were calculated using the provided cross sections and averaging over a Maxwellian velocity distribution using the routine provided by P.S.\,Barklem\footnote{github.com/barklem/libpub}. In a small number of cases, if an insufficiently small energy range was provided to calculate the required rate coefficients, results from another work were used to supplement the data. Care was taken to select cross sections which were in good agreement at the boundary between the different data. In some cases, cross sections had to be read from graphs, but, when this was compared with published rate coefficients, the rate coefficients obtained in this way were within 30 per cent of the published values. This degree of uncertainty in rate coefficients usually alters the ion populations by less than one or two per cent.

In the case of charge transfer of A$^{+z}$ with H, for example, CT rates (in s$^{-1}$) for the models are given by the relation

\begin{equation}
 R_{\rm CT}(T_{\rm e})~=~\frac{N_{\ion{H}{i}}}{N_{\rm H}}~\frac{N_{\rm H}}{N_{\rm e}}~N_{\rm e}~\alpha_{\rm CT}(T_{\rm e})\,,
 \label{eqn:ctrate}
\end{equation}

\noindent where the first term on the right hand side is the fractional population of \ion{H}{i} relative to the total abundance of H, the second term is the relative proportion of H to the electron number density, $N_{\rm e}$ (in cm$^{-3}$,) and $\alpha_{\rm CT}(T_{\rm e})$ is the rate coefficient (in cm$^{3}$\,s$^{-1}$). Based on the discussion in Sect.\;\ref{sec:intro}, modelling the fractional abundances of \ion{H}{i} will obviously require chromospheric modelling. Since this is beyond the scope of this paper, the fractional abundances with height and temperature were taken from the optically-thick, non-LTE radiative transfer calculations of \cite{avrett2008}. This is a semi-empirical model for the average quiet-Sun chromosphere and TR, optimised to give good agreement with observations for the main lines and continua. The total hydrogen and electron densities with height and temperature were also taken from that work. Checking the level populations of \ion{H}{i} in \textsc{Chianti} at the formation temperatures of the low charge states shows that the fraction of hydrogen atoms in excited states is much less than one per cent. It has, therefore, been assumed that neutral hydrogen is all in the ground state for charge transfer reactions. For CT processes with He, the He abundance relative to H is taken from the photospheric value of \cite{asplund2009} and the ion fractional populations were taken from \textsc{Chianti}. Since these ion populations are calculated in the coronal approximation, their effect on CT in the model was checked against those predicted by the new He model of \cite{delzanna2020} at a few temperatures. Although there were some differences in the fractional populations between the two models, the ion populations of carbon and oxygen in this work were not altered when using either source because of the relatively small rates for CT with He. 

All of the CT rate coefficients in the sources are given in $LS$-coupling, while the collisional-radiative models are built in intermediate coupling, (except for \ion{O}{i}). In their fine structure calculations of charge transfer, \cite{stancil1999o1} show that the total rate coefficients at solar temperatures out of the ground term of \ion{O}{i} to be a weighted average of the fine structure rate coefficients and their statistical weights. Since densities are sufficiently high in the lower solar atmosphere, the ions affected by CT generally have their fine structure levels populated according to statistical weight. Accordingly, the $LS$-coupling rates here were split by statistical weight amongst the levels in the initial term involved in the reaction. Reverse reactions are incorporated into the models if transitions end on the ground and metastable levels. Rate coefficients for the reverse process were obtained from the forward reaction using formula (8) given in \cite{stancil1999si4he}. Double CT is possible, here in the case of He as the perturber, and is included where rates are available.

Listed now are the sources used to obtain the rate coefficients for each element. For carbon, CT ionisation between \ion{C}{i} and \ion{C}{ii} requires both radiative and collisional rate coefficients for temperatures which are relevant for the solar atmosphere. These were provided by \cite{stancil1998c1rt} for the transitions connecting ground states, and the cross sections given in \cite{stancil1998c1cs} were used to obtain rate coefficients for transitions involving metastable levels. For CT from the ground of \ion{C}{iii}, the quantal calculations of \cite{errea2015} were the main source. They were supplemented at low energies by the recommended cross section of \cite{janev1988}, which includes radiative CT and is within 10 per cent of the value at the lowest energy point of \cite{errea2015}. For charge transfer from the $2s\,2p\,^3P$ metastable levels the only available cross sections are from \cite{errea2000}. Charge transfer from both the ground and metastable levels do not end on either the ground or metastable levels of \ion{C}{ii}, and so the reverse, CT ionisation reactions are not included for this and more highly charged ions. \cite{errea2015} is also used for CT from \ion{C}{iv} since it is in better agreement with experiment, but only for the low energy values. \cite{tseng1999} is used for higher energies because \cite{errea2015} state they would be more accurate. \cite{liu2003} is used for \ion{C}{v} CT with H, and \cite{yan2013} for reactions involving He, both single and double electron capture in this case. Reactions involving excited levels are not relevant for the last two ions because neither have metastable levels.

For oxygen, charge transfer and charge transfer ionisation rate coefficients were obtained from \cite{stancil1999o1} for transitions between the ground levels of \ion{O}{i} and \ion{O}{ii}. They do not give transitions involving the metastable levels, but \cite{kimura1997} present cross sections for ionisation from the $2s^2\,2p^4\,^1D$ metastable term. The threshold is much higher, principally because this is an endothermic reaction to the $2s\,2p^3\,^2D$ term of \ion{O}{ii}. \cite{barragan2006} provides the rate coefficients for CT from \ion{O}{iii}. \cite{honvault1995} shows that CT ends on the $2s\,2p^4\,^2D,^4P$ terms, and so no reverse reaction has been added. Rate coefficients for CT with H for \ion{O}{iv} come from \cite{wang2003o4}. Their work focuses on charge transfer from the ground term, but they state that a provisional calculation shows that the rate coefficients for the metastable term are only slightly lower than those for the ground. Consequently, the ground term rates have also been used for the metastable term. For CT with He the results of \cite{wu2009o4he} are included. The latter also made available rate coefficients for double CT to \ion{O}{ii}, which have been incorporated, including the reverse reactions. Finally, the rate coefficients provided by \cite{kingdon1996} are used for \ion{O}{v}, which are derived from the calculations of \cite{butler1980}, although these stop at a much lower temperature than where \ion{O}{v} forms.

Calculations for excitation between levels in the same ion through electron capture are more infrequent. During the survey of literature for this work, one process was found, in \cite{errea2000}, for excitation between the ground and metastable term in \ion{C}{iii}. The cross section is an order of magnitude lower than charge transfer and the threshold energy is an order of magnitude higher. This means the process will be much weaker than charge transfer reactions, and more so compared to EIE. This seems to be the agreement generally in the literature about this process, and was not explored further.

\subsubsection{Collisions between neutral atoms and neutral hydrogen}
\label{sec:hcollisions}

Calculations for excitation of neutral atoms by neutral hydrogen are sparse, and rate coefficients are often included using the approximation of \cite{drawin1968}. Later works claim these can overestimate rates by three orders of magnitude. Recently, \cite{barklem2016} developed a new method, and the rate coefficients calculated for neutral oxygen in \cite{barklem2018} were included to test whether this process could influence UV line formation. Rate coefficients were recalculated by P.S.\;Barklem to cover the whole range of \ion{O}{i} formation. \cite{barklem2018} clearly states that the methods used are still approximate and represent an order of magnitude estimate for the process. For collisions involving the excited states of neutral hydrogen, the level populations from the independent atom model of \textsc{Chianti} for hydrogen were used. The rate coefficients for these collisions are already very much lower, however. The only process from these calculations likely to alter the ion balance is charge transfer from the ground and metastable levels of \ion{O}{ii} with H$^-$. The rate coefficients are relatively strong for this, but the rate will depend on the relative abundance of H$^-$. This is highly uncertain, depending on many other conditions in the atmosphere. In this work, it is determined using equations\;(44)\,and\;(54) from the model atmosphere of \cite{vernazza1981}. A departure coefficient of five was chosen, which is the average of the values they give at temperatures relevant to this study.

\section{Results}
\label{sec:results}

In this section, the results of the CR modelling will be presented. The models were all run using a constant pressure of 3$\times$10$^{14}$\,cm$^{-3}$\,K to mimic conditions in the quiet Sun. This is confirmed by the model atmosphere of \cite{avrett2008}, and by \cite{warren2005} from the ratios of density-sensitive lines in the quiet Sun. For the rest of the work, the results using the modelling of \textsc{Chianti} will be referred as the coronal approximation; the CR models which only include the processes given in Sect.\;\ref{sec:elecmethods} will be referred to as electron collisional models; and, those which include all the processes described in Sect.\;\ref{sec:methods} will be called full models. A list of ionisation and recombination rates for the low charge states for each atomic process are given in Tabs\;\ref{tab:ionrates}\;and\;\ref{tab:recrates}. The rates are given at temperatures just above the peak formation temperature of each ion, where ionisation out of the ion begins to have an effect against recombination into the ion from the charge state above.

\subsection{Ion formation}
\label{sec:ionresults}

\subsubsection{Carbon}
\label{sec:cresults}

\begin{figure}
	\centering
	\includegraphics[width=8.4cm]{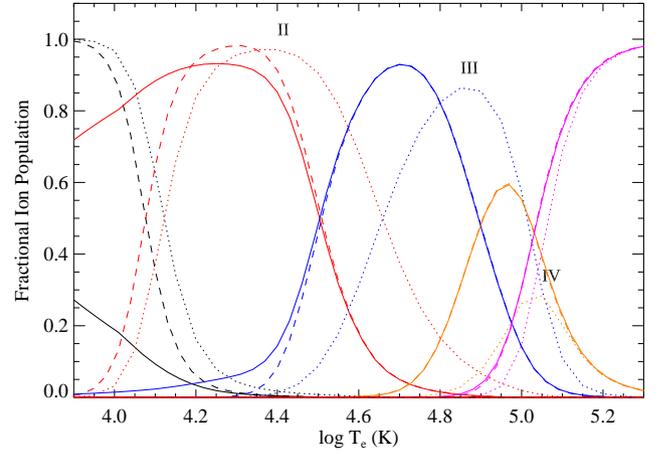}
	\caption[width=1.0\linewidth]{Ionisation equilibrium of carbon: solid line - full model, dashed - Dufresne \& Del Zanna electron collisional model, dotted - \textsc{Chianti} v.9. Ions are highlighted by Roman numerals and different colours.}
	\label{fig:ccrmlrpict}
\end{figure}

On its own, ionisation of \ion{C}{i} by charge transfer would make a minor impact on the electron collisional model, as can be seen from a comparison of the rates in Tab.\;\ref{tab:ionrates}; the process would reduce the \ion{C}{i} population by one or two per cent below 10000\,K. Surveying the CT rates for carbon in both Tab.\;\ref{tab:ionrates}\;and\;\ref{tab:recrates} shows that CT otherwise has no influence on the ion balance. This confirms the conclusions of \cite{maggi2000}, who find that charge transfer of carbon with deuterium in steady state equilibrium has no effect on the ion balance and 30-40\;per cent effect at most on the spectral emission from a tokamak divertor plasma. By contrast, as shown in \cite{dufresne2019}, photo-ionisation is a noticeable effect on the populations in the chromosphere and transition region. The result of the modelling with all the processes modelled in this work is shown in Fig.\;\ref{fig:ccrmlrpict}. The continuum around the threshold of \ion{C}{i} (1101\,\AA) is reasonably strong, and so, when integrating the cross section with the radiance, it makes a significant contribution to the PI rate. In addition, the resonance lines of \ion{C}{iii} (977\,\AA), \ion{C}{ii} (1036\,\AA) and \ion{O}{vi} (1031,\,1037\,\AA), plus the \ion{H}{i} Lyman-$\beta$ 1025\,\AA~line together contribute 20 per cent of the PI rate. The threshold of \ion{C}{ii} (508\,\AA) is slightly longer in wavelength than the threshold of \ion{He}{i} at 504\,\AA, and so the continuum below that wavelength will account for a significant proportion of the PI of \ion{C}{ii}. The cross section, however, is three times smaller than \ion{C}{i} at their respective thresholds. Consequently, PI is much weaker for \ion{C}{ii}, and a smaller population is photo-ionised by comparison. The \ion{He}{ii} 304\,\AA~line contributes 42 per cent to the PI rate.

\begin{table}
	\caption{Comparison of total ionisation rates (in s$^{-1}$) for each process from the ground level of low charge states at the indicated temperature, $T_{\rm e}$ (in K).} 
	\centering	
		\begin{tabular}{p{0.1in}ccccc}
			\hline\hline \noalign{\smallskip}
			Ion & $log\,T_{\rm e}$ & CT & PI & EII \\
			\noalign{\smallskip}\hline\noalign{\smallskip}
			
			\ion{C}{i} & 4.0 & $1.1\times10^{-4}$ & $5.4\times10^{-2}$ & $2.2\times10^{-3}$ \\
			\ion{C}{ii} & 4.4 & $6.1\times10^{-9}$ & $3.8\times10^{-3}$ & $3.2\times10^{-3}$ \\
			\ion{C}{iii} & 4.9 & - & $6.9\times10^{-4}$ & $2.6\times10^{-2}$ \\
			\noalign{\smallskip}
			\ion{O}{i} & 4.0 & 43 & $2.5\times10^{-2}$ & $1.3\times10^{-5}$ \\
			\ion{O}{ii} & 4.5 & - & $9.4\times10^{-3}$ & $1.7\times10^{-4}$ \\
            \ion{O}{iii} & 4.9 & - & $1.3\times10^{-3}$ & $1.3\times10^{-2}$ \\
            \noalign{\smallskip}\hline
		\end{tabular}
	\label{tab:ionrates}
\end{table}

\begin{table}
	\caption{Comparison of total recombination rates (in s$^{-1}$) for each process into the indicated ion from the ground level of the next higher charge state at the given temperature, $T_{\rm e}$ (in K). DR rates include the estimate of suppression due to density.}
	\centering	
		\begin{tabular}{p{0.1in}ccccc}
			\hline\hline \noalign{\smallskip}
			Ion & $log\,T_{\rm e}$ & CT & RR & DR \\
			\noalign{\smallskip}\hline\noalign{\smallskip}
			
			\ion{C}{i} & 4.0 & $7.7\times10^{-7}$ & $1.4\times10^{-2}$ & $1.1\times10^{-2}$ \\
			\ion{C}{ii} & 4.4 & $8.1\times10^{-4}$ & $1.6\times10^{-2}$ & $3.6\times10^{-2}$ \\
			\ion{C}{iii} & 4.9 & $5.3\times10^{-4}$ & $4.3\times10^{-3}$ & $5.0\times10^{-2}$ \\
			\noalign{\smallskip}
			\ion{O}{i} & 4.0 & 21 & $8.2\times10^{-3}$ & $2.3\times10^{-3}$ \\
			\ion{O}{ii} & 4.5 & 0.18 & $9.5\times10^{-3}$ & $9.6\times10^{-3}$ \\
            \ion{O}{iii} & 4.9 & $7.7\times10^{-4}$ & $5.6\times10^{-3}$ & $5.2\times10^{-2}$ \\
            \noalign{\smallskip}\hline
		\end{tabular}
	\label{tab:recrates}
\end{table}

Because the \ion{C}{ii} populations extend to much lower temperatures, this will enhance lines which are formed lower down in the atmosphere. For lines formed at higher temperature, the \ion{C}{ii} populations are reduced at their peak by PI and shifted to lower temperature by density effects on electron collisional processes, which should produce lower intensities for those lines. Arising from the photo-ionisation of \ion{C}{ii} is a population of \ion{C}{iii} which extends far down into the chromosphere. The changes in the formation of the other carbon ions are negligible compared to the level-resolved model without these effects, such that their emission will be unaffected compared to the electron collisional model.

\subsubsection{Oxygen}
\label{sec:oresults}

\begin{figure}
	\centering
	\includegraphics[width=8.4cm]{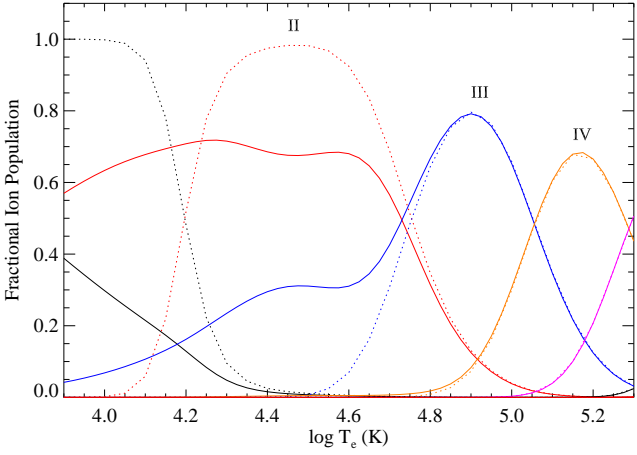}
	\caption[width=1.0\linewidth]{Coronal approximation for oxygen: solid line - this work including photo-ionisation only, dotted - \textsc{Chianti} v.9. Ions are highlighted by Roman numerals and different colours.}
	\label{fig:ocrmpi}
\end{figure}

Considering the influence of PI alone on the charge state distribution, the first ionisation potential of oxygen is similar to carbon, and so the drop in the \ion{O}{i} population is to be expected, given the change seen above. The threshold is 910\,\AA, right below the Lyman-$\alpha$~limit, and so a strong continuum contributes towards PI. Given that there are no strong contributions to the rate from lines, compared to \ion{C}{i}, and that the cross section a third less, the PI rate of \ion{O}{i} is just under half that of \ion{C}{i}. What is perhaps more surprising is the substantial effect PI has on \ion{O}{ii}, as seen in Fig.\;\ref{fig:ocrmpi}. The figure shows the coronal approximation with just PI added, that is, with neither CT nor density effects from suppression of DR and metastable levels. The PI rate from \ion{O}{ii} is more than twice that of \ion{C}{ii}, primarily because the cross section is a factor of two larger. The remaining contribution to the greater PI rate is the contribution from the \ion{He}{i} 304\,\AA~line, which accounts for 52 per cent of the rate. The threshold of \ion{O}{ii} is just under 50\,\AA~from this line, which makes almost three times the contribution to the \ion{O}{ii} PI rate than it does to \ion{C}{ii}. The effect of PI produces a substantial change in the \ion{O}{iii} population at lower temperature. \ion{O}{iv} has a slight increase in its population at lower temperature from PI of \ion{O}{iii}.

\begin{figure}
	\centering
	\includegraphics[width=8.4cm]{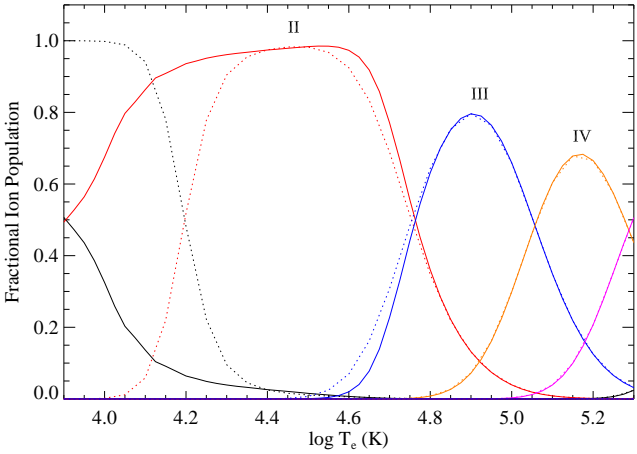}
	\caption[width=1.0\linewidth]{Coronal approximation of oxygen: solid line - this work including charge transfer only, dotted - \textsc{Chianti} v.9. Ions are highlighted by Roman numerals and different colours.}
	\label{fig:ocrmct}
\end{figure}

Because charge transfer between \ion{O}{ii} and \ion{H}{i} is an almost resonant process, the rate coefficients for the forward and reverse reactions are within 20 per cent of each other. CT ionisation begins to dominate just below 10000\,K in the solar atmosphere, at the point where the \ion{H}{ii} population is a little over a half. CT ionisation remains stronger than EII until 10$^5$\,K. CT from \ion{O}{ii} still influences ion formation in the upper chromosphere and lower TR because, without it, there would be no more \ion{O}{i} present. It is also sufficient to maintain a small presence of \ion{O}{i} to higher temperatures than would be present with electron collisional processes alone. Figure\;\ref{fig:ocrmct} illustrates the effect of CT when adding it as a process on its own to the coronal approximation, without PI and density effects on electron collisional processes. It and Tab.\;\ref{tab:recrates} show how, at temperatures close to the peak formation of \ion{O}{ii}, CT is still a strong process as recombination via CT from \ion{O}{iii} also enhances the population of \ion{O}{ii} at higher temperatures than would otherwise be present with electron collisional processes alone. Charge transfer from \ion{O}{iii} is stronger than recombination involving free electrons until $log\,T_{\rm e}=4.8$, above which the normal ion formation curves occur. The strength of this effect is also due to the majority of CT occurring into the $2s\,2p^4\;^4P$ term of \ion{O}{ii}. There is no CT ionisation to inhibit the effect because radiative decays from this term are significantly quicker. Thus, CT has an opposite effect to PI because it reduces the presence of \ion{O}{iii} at lower temperatures, rather than enhancing it. Low temperature lines of \ion{O}{iii}, then, may be a good diagnostic of which processes are affecting ion formation in the solar atmosphere. CT with H is the main cause of CT in the modelling; CT with He contributes just one per cent to the CT rate from \ion{O}{iii} into \ion{O}{ii} and ten per cent to the CT rate from \ion{O}{iv} into \ion{O}{iii}.

\begin{figure}
	\centering
	\includegraphics[width=8.4cm]{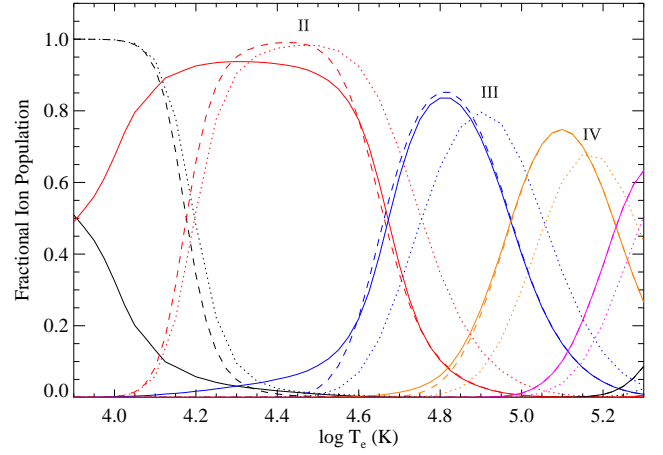}
	\caption[width=1.0\linewidth]{Ionisation equilibrium of oxygen: solid line - full model, dashed - Dufresne et al. electron collisional model, dotted - \textsc{Chianti} v.9. Ions are highlighted by Roman numerals and different colours.}
	\label{fig:ocrmlrpict}
\end{figure}

Because of the magnitude of the rates, it is charge transfer ionisation which governs processes between \ion{O}{i} and \ion{O}{ii} in the chromosphere when all of the processes modelled in this work are included in the final ionisation equilibrium, which is shown in Fig.\;\ref{fig:ocrmlrpict}. Higher up into the atmosphere, at the bottom of the TR, photo-ionisation begins to assert its influence over CT, reducing the peak abundance of \ion{O}{ii} and producing an enhanced population of \ion{O}{iii}. The presence of metastable levels in the modelling contributes to this enhancement of \ion{O}{iii} for the following reason. The CT rate from the metastable $2s^2\,2p^2\;^1D$ term of \ion{O}{iii} is less than half that of the ground term. The metastable term has a population of 18 per cent at TR densities, and so the effective recombination rates into \ion{O}{ii} are reduced. The result is that the \ion{O}{iii} fractional population almost doubles to 0.09 at $log\,T_e=4.5$, compared to when metastable levels are not included in the modelling, and compared to the complete depletion of \ion{O}{iii} when only CT is considered. Clearly, with all the processes included in the CR modelling, \ion{O}{ii} is the dominant ion of oxygen in the upper chromosphere and at the base of the transition region, mainly because of CT. While the \ion{O}{iii} populations are not as strong when the influence of CT is considered together with PI, it still has a presence in this region, which modelling using electron collisional processes alone does not predict.

\subsection{Level populations}
\label{sec:levelresults}

\subsubsection{\ion{O}{i}}

Excitation through collisions with neutral hydrogen was tested on \ion{O}{i} only. At around $7000-10000$\,K in the solar atmosphere the abundance of hydrogen relative to free electrons has dropped significantly compared to its abundance at lower temperatures, according to the model atmosphere of \cite{avrett2008}. The EIE rate coefficients are strong enough at these temperatures that including collisions with neutral hydrogen using the rate coefficients of \cite{barklem2018} does not alter the level populations. It seems this process is more important in the lower chromosphere, such as for the optical lines discussed by \cite{amarsi2018}, where electron collisions are weaker and the hydrogen abundance relative to electrons is substantially higher.

For photo-excitation, the energy separation of the ground and metastable levels is in the optical wavelength range, where the solar spectrum is at its strongest. The PE rate is also dependent on the fifth power of the wavelength, as shown in equation\;(\ref{eqn:perate}). However, the rates are proportional to the radiative decay rates, and these between the ground and metastable levels are very small; at least ten orders of magnitude lower than the dipole decays in \ion{O}{i}, for example. This means PE does not affect the ground and metastable level populations. Furthermore, cascades down to the ground and metastable levels following PE are not sufficient to change the balance of the long-lived levels either. Therefore, PE does not affect ion fractional populations.

\begin{figure}
	\centering
	\includegraphics[width=8.4cm]{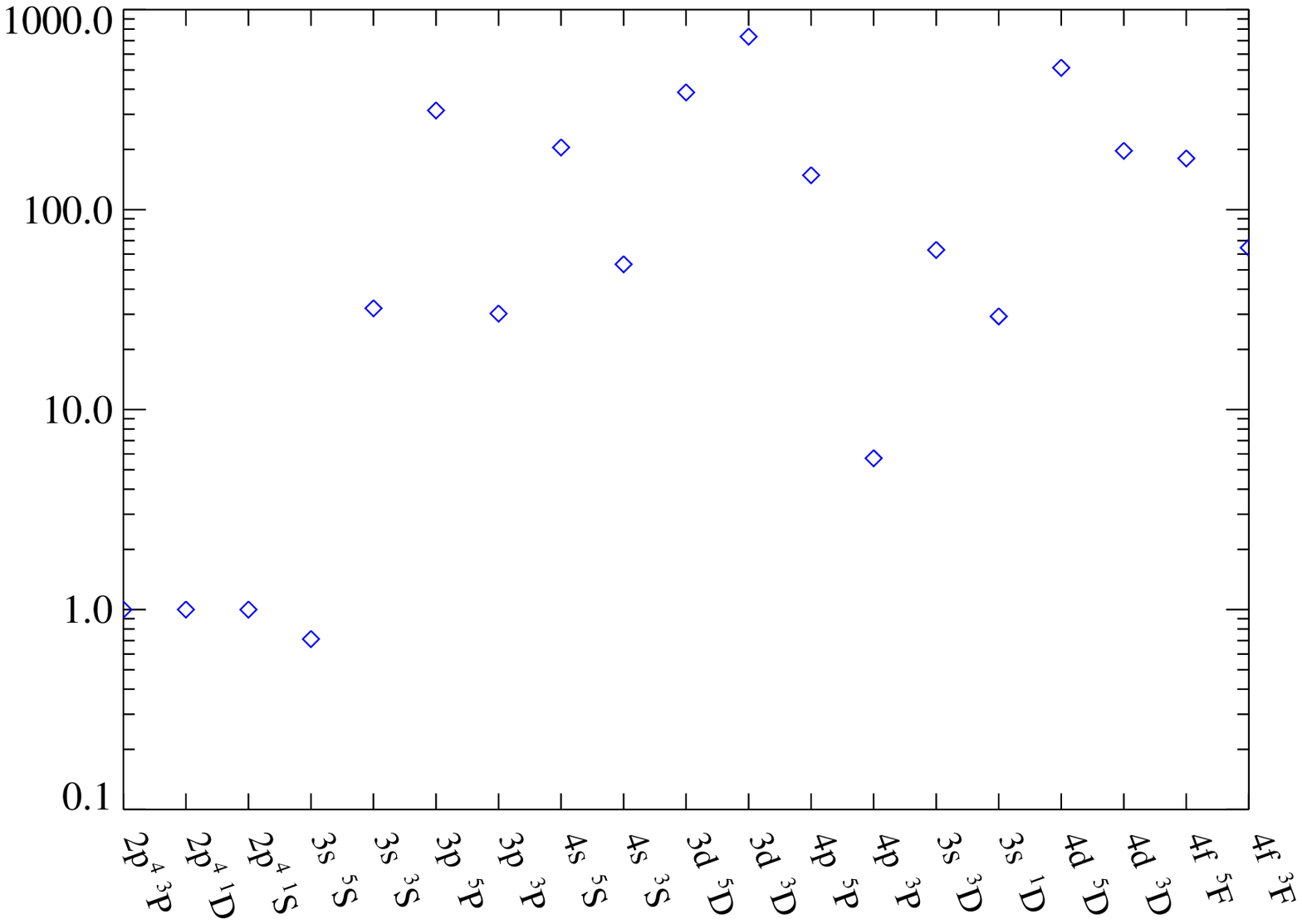}
	\caption[width=1.0\linewidth]{Ratio of \ion{O}{i} term populations from the full model when PE is included to the full model populations when PE is not included, at 10000\,K.}
	\label{fig:olevelpops}
\end{figure}

The excited level populations, however, are substantially altered by PE. Since Bowen fluorescence was first discussed for \ion{H}{ii} regions \citep{bowen1947}, there has been much discussion by a number of authors of how the \ion{H}{i} Lyman-$\beta$ line at 1025.7\,\AA~photo-excites the ground in \ion{O}{i} to the $3d\;^3D$ term. This influences the formation  of the resonant lines around 1304\,\AA~and the intercombination lines at 1355\;and\;1358\,\AA~in the solar setting, (see, for example, \citealt{skelton1982} and \citealt{carlsson1993}). The PE rate from Bowen fluorescence is more than $10^3$ times stronger than the EIE rate to the $3d\;^3D$ term at 10000\,K. When taking into consideration the relative population of the ground term, it makes it the strongest populating mechanism out of all the PE transitions. PE from the ground also produces substantial enhancement to the $3s\;^3D$ and $4d\;^3D$ terms, as shown in Fig.\;\ref{fig:olevelpops}, which shows the ratio of the populations of the terms in the full model with PE included to the populations  in the full model when PE is not included. Following PE from the Lyman-$\beta$ line, downward radiative cascades from $3d\;^3D$ predominantly occur to the $3p\;^3P$ and $3s\;^3S$ terms. There is also some redistribution to the other short-lived terms below through de-excitation by electron collisions, as discussed by the earlier works mentioned in this section. The $3s\;^3S$ term produces the resonance lines around 1304\,\AA, and a factor of 30 increase is seen in its population, which is consistent with the factor of 20 noted by \cite{skelton1982}, who used more approximate atomic data but included radiative transfer.

Discussion has been made in \cite{skelton1982} and \cite{carlsson1993}, for example, about how the Lyman-$\beta$ excitation mechanism enhances the $3s\;^5S$ term, and the corresponding intercombination transitions at 1355.6\,\AA~and 1358.5\,\AA. It does not appear, however, that PE through other transitions has been taken into account in these works and more recently in \cite{lin2015}. The $3s\;^5S$ term has a fractional population of $10^{-7}$ established through EIE, which is relatively large in comparison to the other short-lived levels. Photo-excitation from the ground and metastable terms to this term is negligible, but redistribution of the population of this term through PE affects many of the short-lived terms higher up. The $3p\;^5P$ term is connected to this one by a dipole transition, and the PE rate is of order $10^6$\,s$^{-1}$, while the EIE rate is about $10^3$\,s$^{-1}$. This is caused by a photon with wavelength 7777\,\AA, right at the peak in the solar radiance, and the upper term is enhanced by well over 100 compared to when PE is not included in the model. Similarly, there are dipole transitions from the $3p\;^5P$ term to the $4s\;^5S$ and $3d\;^5D$ terms, and so what ensues is a step-wise excitation process induced by PE. As a result of all of this, the $3s\;^5S$ term is actually depleted by 30 per cent when PE is taken into account for all possible transitions between the 19 terms included in the model. The earlier works mentioned above have noted that predictions for the intercombination lines from the $3s\;^5S$ term are significantly lower than observations, and they did not include all possible channels for PE. Including this effect could further complicate the predictions for these lines. It is noted in passing, in line with the earlier works cited, that photo-ionisation alters the level populations very little, interrupting the very highest levels by only a few per cent.

Overall, the enhancement to the quintet terms is stronger than that to the triplet and quintet systems caused by the Lyman-$\beta$ line. The very highest terms still show marked enhancement, and it suggests that the upward redistribution of population would continue to higher terms, until eventually being suppressed by PI into the continuum. Further enhancement to these levels may be seen through cascades following recombination, as discussed by \cite{lin2015}. It is difficult to know how strong this will be compared to PE, especially because recombination from \ion{O}{ii} is dominated by CT, which occurs between ground levels of oxygen and hydrogen. It is four orders of magnitude stronger than DR and RR. In addition, radiative recombination, which is stronger than DR for this ion at its formation temperature, tends to recombine into the ground complex more than the upper levels. Detailed modelling would be required to determine how enhancement following recombination compares with the effects from photo-excitation seen here.

\subsubsection{\ion{C}{i}}

\begin{figure}
	\centering
	\includegraphics[width=8.4cm]{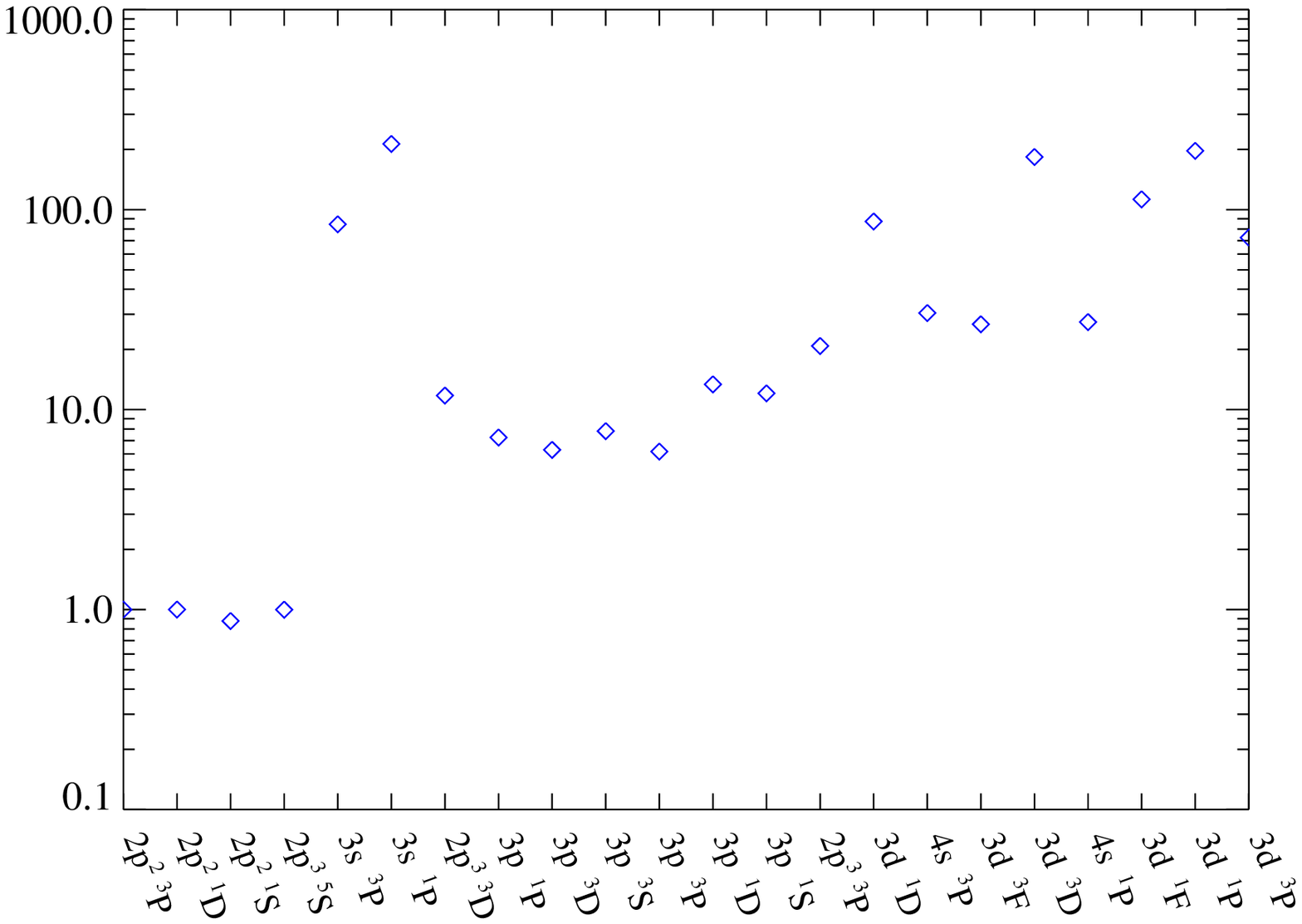}
	\caption[width=1.0\linewidth]{Ratio of \ion{C}{i} term populations from the full model when PE is included to the full model populations when PE is not included, at 7900\,K.}
	\label{fig:clevelpops}
\end{figure}

A similar effect from PE is seen also for \ion{C}{i}, although there are differences in the channels through which the enhancement occurs. For this ion, the enhancement in the triplet terms arises through PE from the ground; the only terms not substantially affected by this are the $3p\;^3D,\,^3P$ terms. The mechanism is responsible for producing the enhancement all the way up to the very highest terms. The first triplet, $3s\;^3P$, above the metastable terms shows a very strong enhancement, but, unlike \ion{O}{i}, there is not a step-wise excitation, as seen in Fig.~\ref{fig:clevelpops}, which shows a plateau in the triplet terms arising from the $2p^3$ and $3p$ configurations. Similarly, this occurs for the singlet states, where the strongest PE comes from the metastable $2p^2\;^1D$ state, with some contribution from the $2p^2\;^1S$ metastable term. Since the $3s\;^3P$ and $2p^3\;^3D$ terms, responsible for emission of the lines around 1657\,\AA~and 1560\,\AA~respectively, are showing population increases by factors of 10-100, this appears to be an important mechanism in their emission. The effect seen here from PE could be a contributing factor to the emission by \ion{C}{i} from levels up to $n=30$ recorded in \cite{sandlin1986}. Again, EII and photo-ionisation do not inhibit the process in this ion.

\subsubsection{Other ions}

For singly-ionised charge states, at the formation temperature of the \ion{C}{ii} UV lines, around 20000\,K, EIE rates have become sufficiently strong that PE changes level populations by only 10-20 per cent in most cases, while at the \ion{O}{ii} UV line formation temperatures the upper level populations are unchanged. Some of the more highly excited levels in \ion{O}{ii}, starting with the $3d$ levels, show changes of 50 per cent up to a factor of four increase in their populations from PE. For higher charge states of carbon and oxygen there is almost no effect on the level populations, apart from one or two isolated levels. No high-$n$ Rydberg states are included in these models, but the results here show that PE may well be sufficient to alter their populations. This could be important when modelling level-resolved recombination into TR ions.

\subsection{Contribution functions}
\label{sec:contribresults}

The discussion in the previous section focused on the level populations at the peak in the effective temperature of the UV lines. Since photo-excitation is independent of density and temperature, levels could become populated at much lower temperature than assumed with free electron collisions. Emission from these levels could arise from lower in the atmosphere, as long as pressure does not increase dramatically. The model atmosphere of \cite{avrett2008} indicates the pressure used here is similar down to 6700\,K. To predict line intensities and compare with observations, however, requires diagnosing the emission measure of the atmosphere through a set of observations. This would mean using models from the coronal approximation for other elements, which may not be accurate for low charge ions because of the changes seen here. Instead, to give an impression of how line emission could be altered by the included effects, contribution functions for some lines have been calculated. The convention given in \cite{delzanna2018} for the contribution function is given by the expression,

\begin{equation}
 C(N_{\rm e},T_{\rm e},\lambda_{ji})~=~A_{ji}\frac{hc}{4\pi\lambda_{ji}}\frac{N^{+z}_j}{N_{\rm e}N^{+z}}~\frac{N^{+z}}{N(X)}\,,
 \label{eqn:contribfn}
\end{equation}

\noindent where $\frac{N^{+z}_j}{N^{+z}}$ is the population of ions of charge $+z$ in level $j$ relative to the total population of the ion, and $\frac{N^{+z}}{N(X)}$ is the ion population as a fraction of the total population of element $X$.

\begin{figure}
	\centering
	\includegraphics[width=8.4cm]{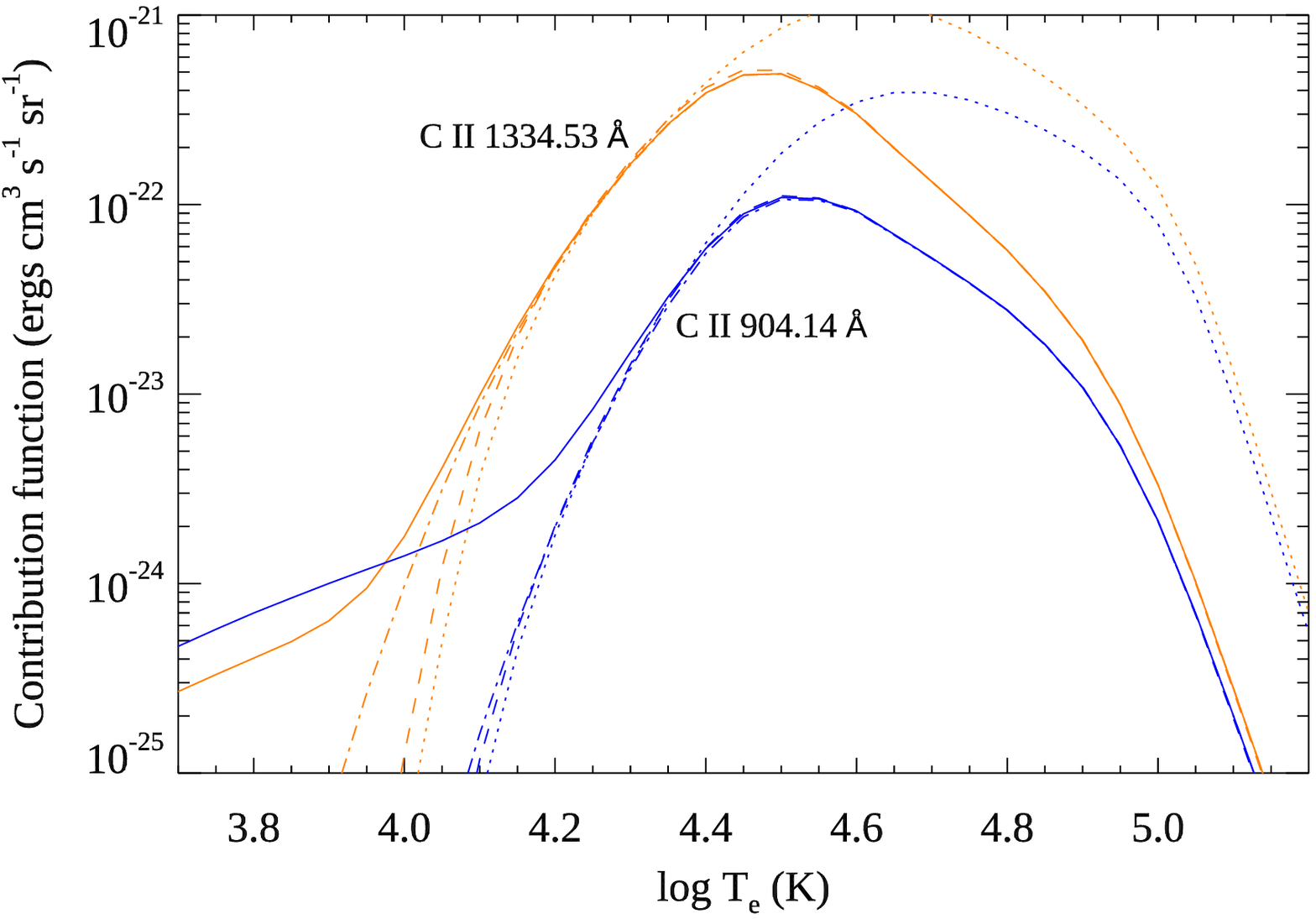}
	\caption[width=1.0\linewidth]{Contribution functions of \ion{C}{ii} 1334.53\,\AA~(orange) and 904.14\,\AA~(blue) lines: solid line - full model including PE, dash-dotted - full model without PE, dashed - electron collisional model of Dufresne \& Del Zanna, and dotted - \textsc{Chianti} v.9.}
	\label{fig:ccontribs}
\end{figure}

\subsubsection{Carbon}

Figure\;\ref{fig:ccontribs} shows the effects from the models on the 1334.53\,\AA~resonance line and 904.14\,\AA~line emitted by \ion{C}{ii}. The decrease in the 1334.53\,\AA~line intensity arising from the shift to lower temperature has already been shown by \cite{dufresne2019} when using the electron collisional model. By the same token, it is evident that a similar decrease in the 904.14\,\AA~predicted intensity would follow when using the electron collisional model. The results from the full model when electron collisions alone are considered as the sole populating mechanism of the upper levels are shown in the figure by the dash-dotted curves. In this case, the resonance line is showing a greater enhancement at lower temperature because PI of \ion{C}{i} allows the presence of \ion{C}{ii} lower in the atmosphere. At the peak in its contribution function, the resonance line is showing a slight decrease compared to the electron collisional model because \ion{C}{ii} populations are decreased due to PI to \ion{C}{iii}. When photo-excitation is included as a populating mechanism for the upper levels, however, as shown by the solid lines in Fig.\;\ref{fig:ccontribs}, the presence of \ion{C}{ii} lower in the atmosphere and the independence of PE to temperature means both lines are quite clearly enhanced at low temperature. Emission may be observed from this ion much lower down than normally assumed. It seems that the 904.14\,\AA~line is showing a slight enhancement from PE right up to the peak in the contribution function because it is not showing the same reduction through PI of \ion{C}{ii} compared to the resonance line. It is obvious that PE is likely to contribute proportionately more to emission of the 904.14\,\AA~line than it would to the 1334.53\,\AA~line. 

\subsubsection{Oxygen}

\begin{figure}
	\centering
	\includegraphics[width=8.4cm]{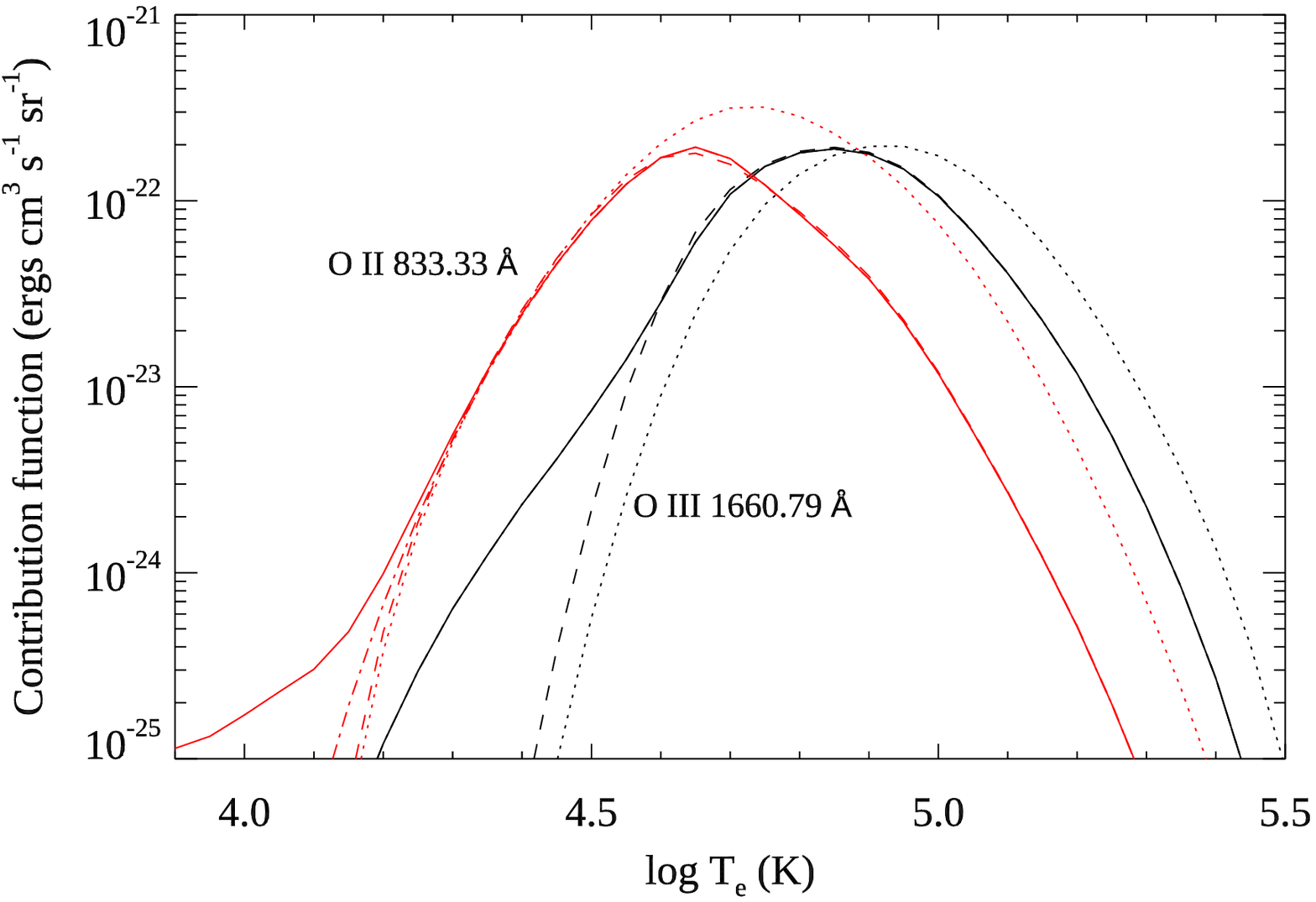}
	\caption[width=1.0\linewidth]{Contribution functions of \ion{O}{ii} 833.33\,\AA~(red) and \ion{O}{iii} 1660.79\,\AA~(black) lines: solid line - full model including PE, dash-dotted - full model without PE, dashed - electron collisional model of Dufresne et al., and dotted - \textsc{Chianti} v.9.}
	\label{fig:ocontribs}
\end{figure}

The changes to the ion fractions is quite clearly reflected in the contribution functions of the oxygen lines shown in Fig.\;\ref{fig:ocontribs}. The decrease in the peak in the contribution function for the 833.33\,\AA~\ion{O}{ii} line with the electron collisional model explains the drop in predicted intensity shown in \cite{dufresne2020}. Compared to that, in the present work, it is noted that the peak rises slightly because charge transfer from \ion{O}{iii} enhances the population of \ion{O}{ii} in the region of $log\,T_{\rm e}=4.6-4.8$. Just below the peak in its contribution function it is lower because of PI of \ion{O}{ii}. When PE is included, it is seen that the line will form at lower temperature because the ion forms below 10000\,K through CT ionisation of \ion{O}{i}. Although the \ion{O}{iii} intercombination line at 1660.79\,\AA~is not formed at all by PE, it is clear that the presence of \ion{O}{iii} much lower in the atmosphere, arising from photo-ionisation of \ion{O}{ii}, allows this line to begin forming lower down. The increase in the predicted intensity of this line shown in \cite{dufresne2020} arises through the higher emission measure present lower in the atmosphere. Given that the enhancement through PI of this line is at lower temperature still and is reasonably close to the peak in the contribution function, it will perhaps further increase the predicted intensity of this line, which was almost a factor of two lower than observations in the earlier work.

The greater contribution to emission from lower in the atmosphere caused by photo-induced processes and charge transfer may alter the profile of the lines. \cite{doschek2004} found that the widths of the intercombination lines were narrower than allowed lines, leading them to offer as a possible explanation that the lines may form in different regions of the Sun. They found the effect is more pronounced for lower charge ions. It is the lower temperature lines, such as the intercombination lines, in the low charge ions which should be affected more by the new processes included here, and may mean their widths are different than higher temperature lines within the same ion. This is not considered by \cite{doschek2004}, but could be a factor contributing to the discrepancy.

\section{Conclusions}
\label{sec:concl}

It is clear that the atomic processes included in this work affect the low charge states of carbon and oxygen significantly. Photo-ionisation causes a depletion in the neutral atoms and a strong enhancement at lower temperatures for singly-charged ions. The peak in the singly-charged ions are also reduced, such that doubly-ionised charge states are present at lower temperatures than would normally be expected. This, along with electron collisional processes moving ion formation to lower temperatures, means that the singly-charged ions predominantly form in the upper chromosphere, rather than in the transition region. Charge transfer has no effect on carbon, but, as discussed in many works, charge transfer ionisation has a strong effect on neutral oxygen. It has been shown here to also affect the next two charge states. It is able to offset some of the depletion of \ion{O}{ii} caused by photo-ionisation, plus it produces a small increase of \ion{O}{ii} at higher temperatures than would normally be present.

Although excitation through collisions with neutral hydrogen has no effect on \ion{O}{i} at the temperatures modelled, its level populations and those of \ion{C}{i} are substantially altered by photo-excitation. Increases by factors of at least ten are seen in the level populations, and in many cases by more than 100 for these two atoms. The excitation continues up to the highest levels included, either directly from the ground and metastable levels in the case of \ion{C}{i}, or in a step-wise manner for \ion{O}{i}. This may be an important contribution to enhancement of the most highly excited levels, in addition to that which may be caused by recombination. Photo-excitation has a limited effect on singly-charged ions of oxygen and carbon, and almost none at all on doubly-charged. In the line contribution functions, the photo-induced processes produce greater enhancement at lower temperature. The presence of an ion at lower temperature because of photo-ionisation, coupled with photo-excitation populating the levels when collisions with free electrons would normally be insufficient, allows emission from lower in the atmosphere. This could have implications for the widths of observed lines and their interpretation should the lines be emitted from regions further down than normally assumed.

The quiet Sun was chosen to model and test the effects of the processes on the charge state distribution because its conditions are those which allow a better assessment of the atomic physics separate from plasma dynamics. In more dynamic environments, such as active regions and flares, it is possible that the effects modelled here will be important for several reasons. Charge transfer rates are likely to scale linearly with electron density, depending on the abundance of hydrogen in the regions, meaning that it should have a similar influence to that shown here. Although PI rates generally dominate more at lower density, if the flux in those regions is enhanced, which will certainly be the case in flares, the PI rates will increase and the process could be consequential. Another environment to which this work may extend is that of prominences. These regions of cooler, higher density plasma in the TR and corona are shown in the recent work of \cite{parenti2019} to have similar pressure to that which is used in the modelling here. The structures are relatively low in the atmosphere, and so the radiation field will be diluted by a similar amount as here. Given the changes seen for the quiet Sun in this work, it indicates that photo-induced processes and charge transfer could play a part in the emission from the low charge states of carbon and oxygen in most regions of the Sun.

\section{Data availability}

In cases where charge transfer and charge transfer ionisation rate coefficients were not given in the original sources, those calculated here from the published cross sections are available at the CDS via anonymous ftp to cdsarc.u-strasbg.fr (130.79.128.5) or via http://cdsarc.u-strasbg.fr/viz-bin/qcat?J/MNRAS. The ion fractional populations at constant pressure derived from the full models, in the \textsc{Chianti} `\texttt{.ioneq}' format, are also available.

\section*{acknowledgements}

	The authors would like to acknowledge the help of P.S. Barklem for clarifying various matters concerning charge transfer, and for producing rate coefficients for the temperature range required. H.E. Mason and P.J. Storey are acknowledged for helpful discussions during the course of the work. The authors appreciate the reviewer for making suggestions to help improve the focus of the work.
	
	Support by STFC (UK) via the Doctoral Training Programme Studentship is acknowledged, plus the support of a University of Cambridge Isaac Newton Studentship. GDZ acknowledges support from STFC (UK) via the consolidated grants to the atomic astrophysics group at DAMTP, University of Cambridge (ST/P000665/1. and ST/T000481/1). NRB is funded by STFC Grant ST/R000743/1 with the University of Strathclyde. \
	
	Most of the atomic rates used in the present study were produced by the UK APAP network, funded by STFC via several grants to the University of Strathclyde. 
	
	\textsc{Chianti} is a collaborative project involving George Mason University, the University of Michigan, the NASA Goddard Space Flight Center (USA) and the University of Cambridge (UK). \

\bibliographystyle{mnras}

\bibliography{pi_co}

\bsp	
\label{lastpage}
\end{document}